\documentclass[aps,prl,reprint,superscriptaddress]{revtex4-1}

\pdfoutput=1 

\usepackage{graphicx}							
\usepackage{hyperref}
\usepackage{float}
\usepackage{amssymb}
\usepackage{amsmath}
\usepackage{siunitx}
\usepackage{tabularx}
\usepackage{bm}
\usepackage{dsfont}					
\usepackage{mathrsfs}				
\usepackage{braket}
\usepackage{mathtools}

\newcommand{\comment}[1]{}			

\DeclareMathOperator{\Tr}{Tr}

\begin{document}

\title{Cavity Quantum Eliashberg Enhancement of Superconductivity}

\author{Jonathan B. Curtis}
\email[]{jcurtis1@umd.edu}
\author{Zachary M. Raines}
\author{Andrew A. Allocca}
\affiliation{Joint Quantum Institute, University of Maryland, College Park, Maryland, 20742, USA}
\affiliation{Condensed Matter Theory Center, University of Maryland, College Park, MD 20742, USA}
\author{Mohammad Hafezi}
\affiliation{Joint Quantum Institute, University of Maryland, College Park, Maryland, 20742, USA}
\author{Victor M. Galitski}
\affiliation{Joint Quantum Institute, University of Maryland, College Park, Maryland, 20742, USA}
\affiliation{Condensed Matter Theory Center, University of Maryland, College Park, MD 20742, USA}
\date{\today}

\begin{abstract}
Driving a conventional superconductor with an appropriately tuned classical electromagnetic field can lead to an enhancement of superconductivity via a redistribution of the quasiparticles into a more favorable non-equilibrium distribution -- a phenomenon known as the Eliashberg effect.
Here we theoretically consider coupling a two-dimensional superconducting film to the quantized electromagnetic modes of a microwave resonator cavity.
As in the classical Eliashberg case, we use a kinetic equation to study the effect of the fluctuating, dynamical electromagnetic field on the Bogoliubov quasiparticles. 
We find that when the photon and quasiparticle systems are out of thermal equilibrium, a redistribution of quasiparticles into a more favorable non-equilibrium steady-state occurs, thereby enhancing superconductivity in the sample.
We predict that by tailoring the cavity environment (e.g. the photon occupation and spectral functions), enhancement can be observed in a variety of parameter regimes, offering a large degree of tunability.
\end{abstract}
\maketitle

It has been known since the late 1960's that subjecting a superconductor to strong microwave radiation can lead to an enhancement of superconductivity~\cite{Wyatt-1966,Dayem-1967}.
The explanation of this was first provided by Eliashberg {\em et. al.}~\cite{Eliashberg-1970, Eliashberg-1971, Eliashberg-1973}, who showed that the irradiation yields a non-thermal distribution of the Bogoliubov excitations with an effectively colder band edge. 
The degree of enhancement can be obtained by using standard BCS theory with a non-thermal quasiparticle distribution function.
In the subsequent decades, Eliashberg's theoretical explanation for this effect has been extended and applied to a variety of other systems~\cite{klapwijk_radiationstimulated_1977,Schmid77,Schmid78, chang_nonequilibrium_1978, Galitski-2009, robertson_dynamic_2011a, Tikhonov-2018}.

In recent years there has been a renewed interest in non-equilibrium superconductivity motivated in-part by a number of ``pump-probe'' experiments which have found that materials subjected to intense $\si{\tera\hertz}$  pulses exhibit transient superconducting properties up to very high sample temperatures~\cite{Cavalleri-2014,Cavalleri-2016,Cavalleri2018}.
Understanding these transient states has led to a variety of theoretical models which go beyond the quasiparticle redistribution effect~\cite{Patel2016, Demler-2017, Kemper-2015, Sentef-2016, Murakami-2016, Komnik-2016}.
   
All of these systems concern the interaction between quantum matter and a {\em classical} external field.
Particularly interesting and novel however, is the effect that a fluctuating {\em quantum} gauge field has on quantum matter.
Indeed, it has been a long-standing focus in the field of cavity-quantum-electrodynamics to realize the dynamical quantum nature of the electromagnetic field through the use of resonant electromagnetic cavities~\cite{purcell,dicke,baskaran,jaynes-cummings,Ebbesen-2016}.
Recently there have been many advances in this area including the realization of exciton-polariton condensates~\cite{exciton-polariton-2002,exciton-polariton-review}, states formed from hybridizing cavity photons and semiconductor excitons.

\begin{figure}
    \centering
    \includegraphics[width=\linewidth]{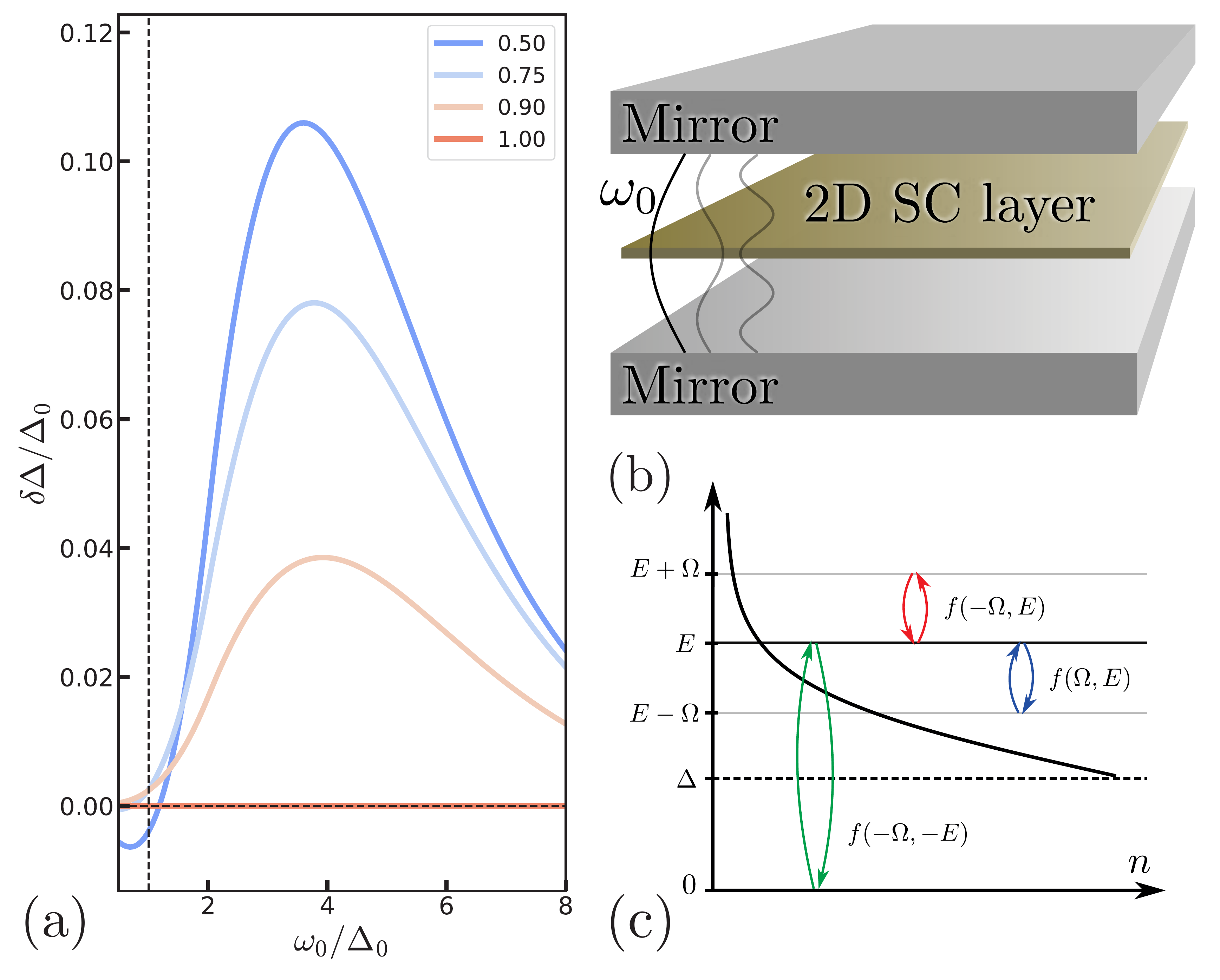}
    \caption{(Color Online) (a) Relative enhancement of the gap function as a function of cavity frequency $\omega_0$ for a particular value of the overall scaling constant $\pi \alpha X D\tau_{\textrm{in}}/c^2$ (we take $X = 133$ and $\pi \alpha D\tau_{\textrm{in}}T_c^2/c^2 = 9.17\times 10^{-5}$ with $T_c$ set to unity). 
    Curves are colored and labeled according to the ratio $T_{\textrm{cav}}/T_{\textrm{qp}}$, comparing the photon and quasiparticle temperatures.
    The enhancement is seen set in after the cavity frequency surpasses the pair-breaking energy $2\Delta_0$.
    (b) Schematic picture of the system used for calculation.
    The lowest cavity resonator mode with cutoff frequency $\omega_0$ is shown, as is the 2D superconducting (SC) layer. 
    (c) Depiction of the various processes which contribute to the quasiparticle collision integral, plotted against the equilibrium $n(E)$.
    The blue arrows depict the down-scattering terms captured by $f(\Omega,E)$, the red arrows depict the up-scattering terms captured by $f(-\Omega,E)$ and the green arrows represent the pair-processes captured by $f(-\Omega,-E)$.
    \label{fig:gap}} 
    \vspace*{-0.2in}
\end{figure}

This paper extends some of these concepts to superconducting systems with an eye on cavity-induced Eliashberg-type enhancement of superconductivity. 
The central observation is that even in a non-equilibrium steady-state the BCS self-consistency equation
\begin{equation}
\label{eq:BCS_sc}
\frac{1}{g} = \int \frac{dE}{E} \nu_{\textrm{qp}}(E) \left[1 - 2n(E)\right]
\end{equation}
can be solved for a non-thermal quasiparticle distribution function $n(E)$, where $\nu_{\textrm{qp}}(E) = 2\nu_F |E|/\sqrt{E^2-\Delta^2}$ is the quasiparticle density of states.
The solution of this equation --  the BCS superconducting gap $\Delta$ -- is therefore a functional of the distribution function $n(E)$ as well as the BCS coupling constant $g$. 
Of particular interest are cases where the gap exceeds its equilibrium thermal value, $\delta\Delta = \Delta[n_F + \delta n] - \Delta[n_F] > 0$. 
In the classical Eliashberg effect, this is achieved via irradiation with a coherent microwave field.
For frequencies smaller than $2\Delta$, pair breaking is suppressed and existing thermal quasiparticles are scattered up to higher energies, where their debilitating effect is lessened by the reduced relative density of states.
This emptying of states near the band edge increases $\Delta$ above its equilibrium value.
In this paper we generalize this idea to include the dynamical fluctuations of the electromagnetic field in a microwave cavity, depicted in the inset of Fig.~\ref{fig:gap}(b).
Our main result is that, by appropriately tuning the parameters of the cavity environment (e.g. resonance, line-width, temperature, etc), an enhancement in the BCS gap strength may be obtained, now in the absence of coherent electromagnetic radiation.
This gap enhancement is shown in Fig.~\ref{fig:gap}(a), which illustrates the change in the BCS gap strength $\delta\Delta$ as a function of the cavity resonant frequency $\omega_0$.
The rest of the paper is devoted to deriving this result.

We begin with a model of an s-wave superconductor described by the BCS Hamiltonian (setting $\hbar = k_B = 1$) 
\begin{equation}
H = \int d^2r \left[\psi^\dagger_\sigma\left(-\frac{\mathbf{D}^2}{2m} - \mu\right)\psi_\sigma
- g \psi^\dagger_\uparrow\psi^\dagger_\downarrow \psi_{\downarrow} \psi_\uparrow
\right],
\label{eq:BCS-Hamiltonian}
\end{equation}
where $\psi_{\sigma}$ is the electron field operator, which is minimally coupled to the electromagnetic vector potential $\mathbf{A}$ through the gauge covariant derivative $\mathbf{D} = \nabla+ie\mathbf{A}$.
Throughout we will employ the radiation gauge $\nabla\cdot\mathbf{A} = 0$.
The interaction is decoupled via standard mean-field theory, and the resulting Hamiltonian is diagonalized with a Bogoliubov transformation
\begin{equation} 
\begin{pmatrix}
\psi_{\mathbf{p},\uparrow}\\
\psi^\dagger_{-\mathbf{p},\downarrow}
\end{pmatrix} 
= 
\begin{pmatrix}
u_\mathbf{p} & -v_\mathbf{p}\\
v_\mathbf{p} & u_\mathbf{p}\\
\end{pmatrix}
\begin{pmatrix}
\gamma_{\mathbf{p},+} \\
\gamma^\dagger_{-\mathbf{p},-}\\ 
\end{pmatrix},\ 
u, v = \sqrt{\frac{1}{2}\left(1 \pm \frac{\xi}{E}\right)},
\label{eq:gammas}
\end{equation}
where $\gamma_{\mathbf{p}\pm}$ are the Bogoliubov quasiparticle (BQP) annihilation operators, $E_\mathbf{p} = \sqrt{\xi^2_\mathbf{p} + \Delta^2}$ is the BQP dispersion, and $\xi_{\mathbf{p}}=\mathbf{p}^2/2m - \mu$.
The electromagnetic field $\mathbf{A}$ is subject to cavity quantization of the transverse-momentum, leading to a dispersion relation for in-plane momentum $\mathbf{q}$ of 
\begin{equation} 
\omega_{n,\mathbf{q}} =  \sqrt{{\left(\frac{n \pi c}{L}\right)}^2 + c^2\mathbf{q}^2} \equiv \sqrt{n^2\omega_0^2 + c^2\mathbf{q}^2}
\end{equation}
where $n = 1,2,3,...$ indexes the harmonic of the confined mode.
For simplicity, we will only consider the fundamental $n=1$ harmonic and place the superconducting sample at the anti-node where the coupling to the field is strongest, as depicted in Fig.~\ref{fig:gap}(b).

To leading order, the interaction between photons and BQPs obtained from Eq.~\eqref{eq:BCS-Hamiltonian} occurs through the coupling of the vector potential to the electronic current via 
\[
H^{\textrm{int}} = -e\int d^d r\mathbf{j}\cdot\mathbf{A}.
\]
Applying the Bogoliubov transformation and Fourier transforming to momentum space this becomes 
\begin{multline}
\label{eq:paramagnetic}
    \mathbf{j}_{\mathbf{q}} = \int_p \frac{\mathbf{p} - \frac12\mathbf{q}}{m}\left[
    \left(
    u_\mathbf{p-q} u_{\mathbf{p}} + v_{\mathbf{p-q}} v_{\mathbf{p}}
    \right)
    \gamma^\dagger_{\mathbf{p-q},\sigma}\gamma_{\mathbf{p},\sigma}\right.\\
    \left.+
    \left(
    u_\mathbf{p-q} v_{\mathbf{p}} - v_{\mathbf{p-q}} u_{\mathbf{p}}
    \right) \left(
    \gamma^\dagger_{\mathbf{p-q},+}\gamma^\dagger_{-\mathbf{p},-}
    - \gamma_{\mathbf{p},+} \gamma_{-(\mathbf{p-q}),-}
        \right)\right],
\end{multline}
where we use the shorthand $\int_p\dotsi = \int d^2p\dotsi/(2\pi)^2$.
We see there are three types of matrix element appearing in Eq.~\eqref{eq:paramagnetic}, corresponding to scattering (by both emission and absorption of photons), pair-breaking, and pair-recombination respectively.
Through these processes, the fluctuating cavity photon field will induce transitions amongst the BQP eigenstates, resulting in a redistribution of the quasiparticle occupations. 
This is described by a kinetic equation
\begin{equation}
    \frac{\partial n_{\mathbf{p}}}{\partial t}  = 
    \mathcal{I}_{\text{cav}}[n]  - \frac{n_{\mathbf{p}} - n_F\left(\frac{E_{\mathbf{p}}}{T_{\textrm{qp}}} \right) }{\tau_{\text{in}}}.
    \label{eq:kinetic}
\end{equation}
The first term on the RHS describes the photon-induced pairing/de-pairing and scattering of quasiparticles while the second term describes a generic inelastic relaxation mechanism which describes the coupling to a phonon bath at temperature $T_{\textrm{qp}}$.
The approximation here is that the inelastic relaxation rate $\tau_{\textrm{in}}^{-1}$ is small compared to other energy scales, as was assumed in the original work of Eliashberg~\cite{Eliashberg-1973,klapwijk_radiationstimulated_1977,Schmid77}.

In this limit we can perturbatively solve for the steady-state of the kinetic equation~\eqref{eq:kinetic} by expanding in small deviations $\delta n = n - n_F$ from equilibrium.
To lowest order, the correction is $\delta n = \tau_{\text{in}}\mathcal{I}_{\text{cav}}[n_F] $.
Utilizing the detailed balance properties of thermal equilibrium, this will end up depending on the photon occupation function $N(\omega)$ through its deviation from equilibrium:  
\begin{equation}
    \label{eq:photon-dist}
    \delta N_{\textrm{cav}}(\omega) \equiv N(\omega) - n_B\left(\frac{\omega}{T_{\textrm{qp}}}\right),
\end{equation}
where $n_B(z)$ is the Bose occupation function.

To compute the cavity-induced collision integral, we rely on Fermi's Golden Rule (FGR), applied to both the pairing/de-pairing and scattering processes. 
The result is 
\begin{multline}
\label{eqn:collision}
\mathcal{I}_{\textrm{cav}}[n] = \int_{p'}\bigg\{ 
\Gamma^{\textrm{pair}}_{\mathbf{p},-\mathbf{p}'} \Big[(1-n_{\mathbf{p}}) (1-n_{-\mathbf{p}'}) N(E_{\mathbf{p}} + E_{-\mathbf{p}'}) \\ 
- \left(n_{\mathbf{p}} n_{-\mathbf{p}'} \left(N(E_{\mathbf{p}} + E_{-\mathbf{p}'}) + 1 \right)\right)\Big] \\
+\bigg(\Gamma^{\textrm{scat}}_{\mathbf{p}'\to\mathbf{p}}\Big[ n_{\mathbf{p}'} \left(1-n_{\mathbf{p}}\right)\left( N(E_{\mathbf{p}'}-E_{\mathbf{p}})+1\right)\\
    - \left(1- n_{\mathbf{p}'}\right) n_{\mathbf{p}}N(E_{\mathbf{p}'}-E_{\mathbf{p}}) \Big] - \left(\mathbf{p}\leftrightarrow \mathbf{p}'\right)\bigg)\bigg\}\\
\end{multline}
with the $\Gamma$'s given by  
\begin{gather}
\begin{multlined}
\label{eq:pair-rate}
\Gamma^{\textrm{pair}}_{\mathbf{p},-\mathbf{p}'} =  \frac{e^2}{2\epsilon_0\omega_{\mathbf{p-\mathbf{p}'}}} \sum_{\alpha}\bigg|\bm{\epsilon}_{\alpha,\mathbf{p}-\mathbf{p}'} \cdot\left(\frac{\mathbf{p}+\mathbf{p}'}{2m}\right)\bigg|^2\\
\times\left(u_{\mathbf{p}}v_{-\mathbf{p}'} - u_{-\mathbf{p}'}v_{\mathbf{p}}\right)^2 \mathcal{A}_{\mathbf{p}-\mathbf{p}'}\left(E_{\mathbf{p}}+E_{-\mathbf{p}'}\right)
\end{multlined}\\
\begin{multlined}
    \label{eq:scat-rate}
    \Gamma_{\mathbf{p}\rightarrow\mathbf{p}'}^{\textrm{scat}} =  \frac{e^2}{2\epsilon_0\omega_{\mathbf{p-\mathbf{p}'}}} \sum_{\alpha}\bigg|\bm{\epsilon}_{\alpha,\mathbf{p}-\mathbf{p}'} \cdot\left(\frac{\mathbf{p}+\mathbf{p}'}{2m}\right)\bigg|^2\\
\times\left(u_{\mathbf{p}}u_{\mathbf{p}'} + v_{\mathbf{p}'}v_{\mathbf{p}}\right)^2 \mathcal{A}_{\mathbf{p}-\mathbf{p}'}\left(E_{\mathbf{p}}-E_{\mathbf{p}'}\right).
\end{multlined}
\end{gather}
These contain the dependence on the cavity mode polarization vectors $\bm{\epsilon}_{\alpha\mathbf{q}}(z=L/2)$, the (momentum resolved) photon spectral function  
\begin{equation}
\label{eq:spectral}
\mathcal{A}_\mathbf{q}(\omega) = \frac{1/\tau_{\textrm{cav}}}{ (\omega - \omega_{\mathbf{q}})^2 + (1/2\tau_{\textrm{cav}})^2 },
\end{equation}
with photon lifetime $\tau_{\textrm{cav}}$, and the squares of BCS coherence factors
\begin{gather}
\label{eq:coherence}
(u_\mathbf{p} v_{-\mathbf{p'}} - v_{-\mathbf{p'}} u_\mathbf{p})^2 = \frac{1}{2}\left(1 - \frac{\xi_{\mathbf{p}} \xi_{-\mathbf{p}'}+ \Delta^2}{E_{\mathbf{p}} E_{-\mathbf{p}'}}\right)\\
(u_\mathbf{p}u_\mathbf{p'}+v_\mathbf{p}v_\mathbf{p'})^2 = \frac{1}{2}\left(1 + \frac{\xi_\mathbf{p} \xi_\mathbf{p'}+ \Delta^2}{E_\mathbf{p} E_\mathbf{p'}}\right).
\end{gather}

These collision integrals are derived based on the assumption of a perfectly clean sample, and so momentum is conserved.
In reality however, impurities are always present in a quasi-two dimensional sample and should not be ignored.
Given that the photons of relevance are of long wavelengths, it is appropriate to invoke the quasiclassical approximation whereby we restrict our attention to states near the Fermi surface.
In the limit of strong disorder (as compared to the gap) we then can incorporate elastic impurity scattering by replacing the photonic momentum-conserving delta function $(2\pi)^2 \delta( \mathbf{q} - (\mathbf{p}-\mathbf{p}'))$ with a constant $(\nu_F/\tau_{\textrm{el}} )^{-1}$, where $\mathbf{q}$ is the momentum transferred to the photon, $\nu_F$ is the density of states per spin at the Fermi level, and $\tau_\textrm{el}$ is the elastic scattering time~\cite{Mattis-Bardeen}.
We are then free to independently perform the integrals over the direction of the momentum. 
The validity of this heuristic may be confirmed by appealing to e.g. the solution of the Usadel equation~\cite{Usadel1970} or the Keldysh non-linear sigma model~\cite{Feigelman-2000,Tikhonov-2018,kamenev}, which describe the quasiclassical collective modes of the strongly disordered superconductor (as described in the supplement~\footnote{See \href{http://link.aps.org/supplemental/10.1103/PhysRevLett.122.167002}{URL} for supplemental material employing the full Non-linear Sigma Model calculation, including reference~\cite{Diehl-2016}}).

The result of this procedure is a collision integral which is a function of the quasiparticle energy only.
Evaluating the correction to the quasiparticle distribution function, we find 
\begin{equation}
\label{eqn:dist-correct}
\delta n(E) = \tau_{\textrm{in}}\int_{-\infty}^{\infty} d\Omega J_{\textrm{cav}}(\Omega)\delta N_{\textrm{cav}}(\Omega) K(\Omega,E),
\end{equation}
where $K(\Omega,E) = f(\Omega,E) + f(-\Omega,E) - f(-\Omega,-E)$, with 
\begin{multline}
f(\Omega,E) = \theta(E-\Omega-\Delta) \frac{\nu_{\textrm{qp}}(E-\Omega)}{\nu_F}\times\\
\frac12 \left( 1+\frac{\Delta^2}{E(E-\Omega)}\right)\left[ n_F\left(\frac{E-\Omega}{T_{\textrm{qp}}}\right) - n_F\left(\frac{E}{T_{\textrm{qp}}}\right)\right].
\end{multline}
Here $\theta(x)$ is the Heaviside step-function.
The three $f$ terms appearing in $K(\Omega,E)$ are depicted schematically in Fig.~\ref{fig:gap}(c), alongside the various processes they describe.
After the Fermi-surface average, the coupling to the cavity is effectively characterized by the coupling function 
\begin{equation}
\label{eq:cavity-function}
J_{\textrm{cav}}(\Omega) = 4\pi \alpha c D \int \frac{d^2\mathbf{q}}{(2\pi)^2} \frac{\mathcal{A}_{\mathbf{q}}(\Omega)}{2\omega_q} \sum_{\alpha}|\hat{\bm{\epsilon}}_{\alpha\mathbf{q},\parallel}|^2,
\end{equation}
where $D = v_F^2 \tau_{\textrm{el}}/2$ is the electronic diffusion constant and $\hat{\bm{\epsilon}}_{\alpha\mathbf{q},\parallel}$ indicates that only the in-plane components of the polarization vector contribute.
For a BCS gap of order $\Delta = \SI{10}{\kelvin}$ we find a corresponding resonance frequency $\omega_0 \sim \SI{1.3}{\THz}$.
Recently, a number of advances have lead to large enhancements in the strength and tunability of the light-matter coupling strength in this frequency regime, such that $J_\textrm{cav}(\Omega)$ may potentially exceed what is expected from our simple planar cavity model by many orders of magnitude~\cite{Blais2004,Maissen2014,Malerba2016,Bayer2017}.
We incorporate this fact by rescaling the spectral function $J$ by a phenomenological factor $X$, so that $J(\Omega) \rightarrow \tilde{J}(\Omega) = X J_{\textrm{cav}}(\Omega)$.

\begin{figure}[tp]
    \includegraphics[width=\linewidth]{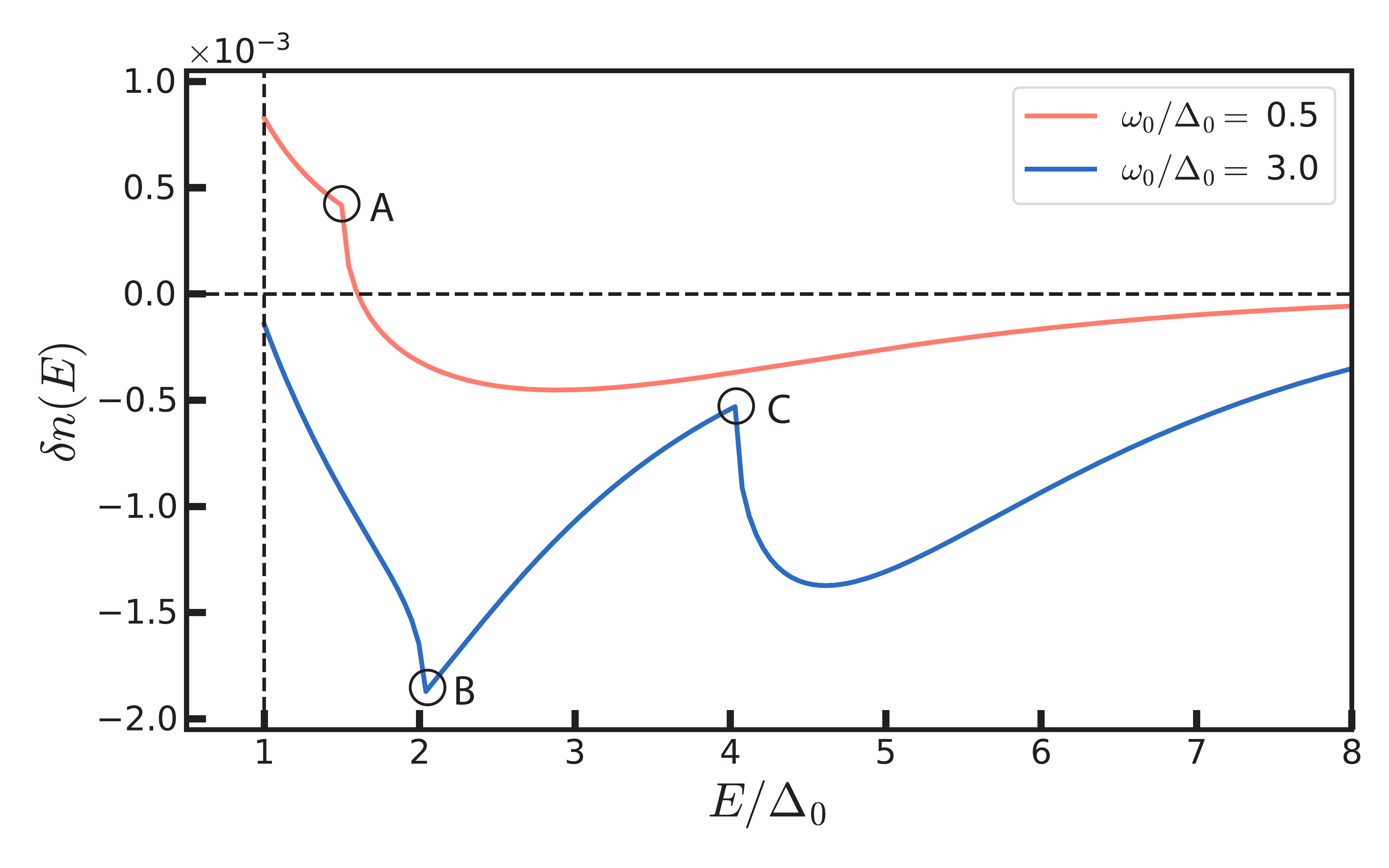}
    \vspace{-24pt}
    \caption{(Color Online) Change in quasiparticle distribution function due to cavity photons.
    The two curves are at the same temperature ($T_{\textrm{cav}}/T_{\textrm{qp}} = 0.5$) but different cavity frequencies $\omega_0/\Delta_0$. 
    For low cavity frequency (orange), the gap $\Delta$ is diminished due to an accumulation of cooler quasiparticles near the gap-edge, due to a down-scattering of particles.
    For higher cavity frequency (blue), the recombination processes are more dominant and lead to a net reduction in quasiparticles, enhancing the gap $\Delta$.
    The kink-features labeled $A$ and $C$ reflect the onset of the term $f(\Omega,E)$ in Eqn.~\eqref{eqn:dist-correct}, which is only non-zero for $E > \omega_0 + \Delta_0$.
    At higher cavity frequencies ($\omega_0>2\Delta_0$) an additional kink-feature (located at $B$) emerges at $E = \omega_0 - \Delta_0$.
    For $E < \omega_0 - \Delta_0$, the term $f(-\Omega, E)$ (which represents the pair-processes) contributes over the entire integration region of $\Omega>\omega_0$, while for $E>\omega_0 - \Delta_0$ the integral only captures some of the frequencies where this term contributes.
    \label{fig:distribution}}
    \vspace*{-6mm}
\end{figure}

In order to simplify the calculation, we will study the system in the Ginzburg-Landau (GL) regime ($T_{\textrm{qp}} \lesssim T_{c}$), which allows us to expand the gap equation in powers of $\Delta$.
Including the non-equilibrium distribution function contribution, this results in
\begin{equation}
\left( \frac{T_c-T_{\textrm{qp}}}{T_c} - \frac{7\zeta(3)}{8\pi^2} \frac{\Delta^2}{T_c^2} -2 \int_{\Delta}^{\infty} \frac{dE}{E}\frac{\nu_{\textrm{qp}}(E)}{\nu_F} \delta n(E)  \right)  \Delta = 0. 
\end{equation}
To leading order in the gap change, we obtain the correction to the BCS gap 
\begin{equation}
\frac{\delta \Delta}{\Delta_0} = -\frac{T_c}{T_c - T_{\textrm{qp}}} \int_{\Delta_0}^{\infty} \frac{dE}{E}\frac{\nu_{\textrm{qp}}(E)}{\nu_F} \delta n(E).
\end{equation}
This is plotted in Fig.~\ref{fig:gap}(a) as a function of the cavity frequency $\omega_0$ for different photon temperatures relative to the quasiparticle temperature $T_{\textrm{qp}}$.
The enhancement is ultimately driven by the enhanced BQP recombination rate which, for a cold photon reservoir serves to remove detrimental quasiparticles.

This can be seen explicitly in Fig.~\ref{fig:distribution}, which shows the change in the distribution function $\delta n$ for two different cavity frequencies.
When the cavity frequency is too low, scattering-processes dominate and the photons cool the existing BQPs, leading to a build-up of particles near the gap edge.
At higher cavity frequencies the pair-processes dominate, leading to an enhancement as photons now cool the system by reducing the total number of harmful BQPs.

\begin{figure}
    \centering
    \includegraphics[width=\linewidth]{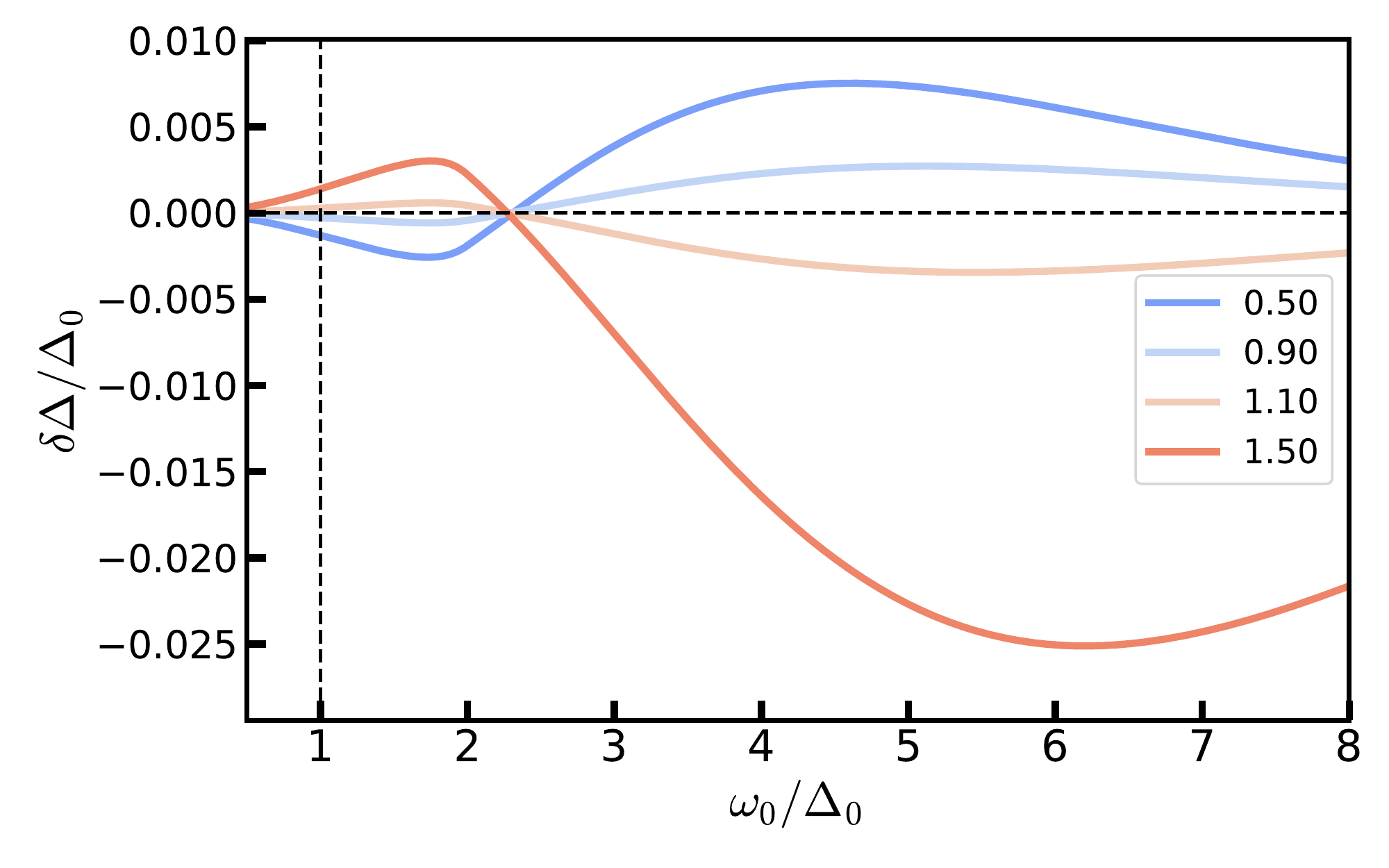}
    \caption{(Color Online) Gap enhancement $\delta \Delta_0$ for a single-mode cavity, for both cold and hot photons.
    The y-axis is determined by the overall scale $4\pi \alpha D \tau_{\textrm{in}} T_c^2/((\pi \sqrt{3})^3 c^2) X$; with the same values chosen for $X$ and $\tau_{\textrm{in}},\tau_{\textrm{el}},v_F/c$ as in Fig.~\ref{fig:gap}. 
    Curves are colored and labeled according to the ratio $T_{\textrm{cav}}/T_{\textrm{qp}}$, comparing the photon and quasiparticle temperatures.
    Here the cavity width is held fixed at $1/2\tau_{\textrm{cav}} = 10 \omega_0$.}
    \label{fig:SM_gap}
\end{figure}

While the effect we predict here essentially relies on the cooling ability of the cold photon reservoir, we also remark that our formula for $\delta n(E)$, presented in Eq.~\eqref{eqn:dist-correct}, is valid for a wide-variety of photon spectral functions.
In particular, switching from a multi-mode planar cavity, where $J_{\textrm{cav}}(\Omega)\sim \omega_0 (1+\omega_0^2/\Omega^2)\theta(\Omega-\omega_0)$ is roughly constant for $\Omega > \omega_0$, to a simpler single-mode cavity, where $J_{\textrm{cav}}\sim \omega_0^2 \frac{2\kappa}{(\Omega-\omega_0)^2+\kappa^2}$ is peaked at the resonant frequency, will allow for an enhancement in $\delta \Delta$ even when the photon reservoir is hotter than the sample.
This is explicitly demonstrated in Fig.~\ref{fig:SM_gap}, where we plot $\delta \Delta$ against $\omega_0$ for the case of a single-mode $J_{\textrm{cav}}(\Omega)$.
The enhancement in $\delta \Delta$ due to hot photons is now qualitatively similar to the classical Eliashberg effect, albeit with a narrow spectral broadening applied to the driving.
For cold photons, the enhancement is similar to that seen in the multi-mode system and results from the photons cooling the sample via enhanced BQP recombination. 

\comment{DOUBLE CHECK PRE-FACTOR FOR THIS SET UP, ALSO DOUBLE CHECK Q FACTOR }

In conclusion, we have generalized the classical Eliashberg effect to include both quantum and thermal fluctuations, as realized by a thermal microwave resonator cavity.
In the appropriate parameter regime, we show that the photonic reservoir can be used to drive the quasiparticles into a non-equilibrium state which enhances the superconducting gap $\Delta$.
In our calculation, we assumed that the cavity relaxation rate $\tau_{\textrm{cav}}^{-1}$ was fast, allowing us to essentially ignore the dynamics and kinetics of the photons themselves.
We should not expect this to remain the case when we go to the limit of a high-quality cavity, in which the relaxation rate $\tau_{\textrm{cav}}^{-1}$ is no longer small compared to all the other energy scales in the problem.
In the high-quality limit, a more elaborate treatment which treats the joint evolution of fermion-photon system is required. 
Though potentially much more complicated, the inclusion of photons as a participating dynamical degree of freedom may unveil many new and interesting phenomena.
These range from the formation of new collective modes (including polaritons)~\cite{sentef_cavity_2018,Allocca}, superradiant phases~\cite{baskaran,flatiron}, and potentially photon-mediated superconductivity~\cite{cavalleri-2018-cavity}.
The prospect of exploring the full breadth of these joint matter-gauge systems is an exciting development in the fields of quantum optics and condensed matter physics.

\begin{acknowledgments}
The authors would like to thank Jacob Taylor and Gil Refael for productive discussions.
This work was supported by NSF DMR-1613029 and US-ARO (contract No. W911NF1310172) (Z.R.), the National Science Foundation Graduate Research Fellowship Program under Grant No. DGE1322106 (J.C.), DARPA DRINQS project FP-017, ``Long-term High Temperature Coherence in Driven Superconductors" (A.A.), AFOSR FA9550-16-1-0323, ARO W911NF-15-1-0397, and NSF Physics Frontier Center at the Joint Quantum Institute (M.H.), and DOE-BES (DESC0001911) and the Simons Foundation (V.G.).
\end{acknowledgments}

\bibliography{references}







\clearpage
\onecolumngrid
\appendix*


\section{Keldysh Non-Linear Sigma model}

In order to derive the correction to the quasiparticle distribution functions in the presence of disorder, we employ the Keldysh nonlinear $\sigma$ model (KNL$\sigma$M) as derived by \textcite{Feigelman-2000}.

\subsection{Schematic derivation of the model}
We first briefly outline the derivation of the Keldysh nonlinear sigma model before describing the calculations performed in our work.
For more details on the KLN$\sigma$M we refer the reader to \textcite{Feigelman-2000} or \textcite{kamenev}.

The derivation of the sigma model begins with a minimally coupled BCS action on the Keldysh contour in the presence of a random impurity potential
\begin{equation}
S = \oint_C dt d\mathbf{x} \left[\bar{\psi}\left(i \partial_t - \hat{\epsilon}\left(-i\nabla + \frac{e}{c}\mathbf{A}\right) + \mu - V_\text{imp}\right)\psi
 + \frac{\lambda}{\nu}\bar{\psi}_\uparrow\bar{\psi}_\downarrow \psi_\downarrow \psi_\uparrow\right]
\end{equation}
with $\hat{\epsilon}$ being the quasielectron energy, $\mu$ the chemical potental, $\nu$ the density of states at the Fermi surface, $\lambda$ the BCS coupling strength, $V_\text{imp}$ is the impurity potential.
$\oint_C$ denotes integration over the Keldysh contour.
One now averages over gaussian disorder which induces an effective disorder interaction in the usual manner
\begin{equation}
    iS_\text{dis} = -\frac{1}{4\pi\nu\tau}\int_C dt dt' d\mathbf{x}\bar{\psi}(t)\psi(t) \bar\psi(t')\psi(t').
\end{equation}
The bilinears $\bar{\psi}(t)\psi(t)$ describe rapidly varying modes on the length scales of the impurities.
However, the bilinears $\bar\psi(t)\psi(t')$ describe slowly varying degrees of freedom.
Therefore a Hubbard-Stratonovich field $Q$ dual to $\bar\psi(t)\psi(t')$ is introduced to decouple the disorder interaction.
The BCS interaction is also decoupled via the Hubbard-Stratonovich field $\Delta$ in the usual fashion.
Coupling to the $A$-field is handled via the paramagnetic coupling $\mathbf{j}\cdot \mathbf{A} \approx \frac{e}{c}\mathbf{v}_F \cdot \mathbf{A}$.
At this point one performs the Larkin-Ovchinnikov rotation and integrates out the fermions.
This leads to an action for the Hubbard-Stratonovich fields $Q$ and $\Delta$
\begin{equation}
    iS = -\frac{\pi\nu}{8\tau} \Tr \check Q^2 + \Tr\ln\left[\check G^{-1} + \frac{i}{2\tau}\check Q - \frac{e}{c}\mathbf{v}_F\cdot\check{\mathbf{A}} + \check\Delta\right]
\end{equation}
where $G$ is the Bogoliubov-de Gennes Green's function.
One then performs an expansion about the saddle-point solution for $Q$ as well as a gradient expansion.
One notes that the $\Tr Q^2$ vanishes on the soft manifold $Q^2=1$ \textemdash where we must keep in mind that the unit matrix must have the proper analyticity structure \textemdash indicating that such modes are massless.
The result of these expansions along with the non-linear constraint gives the KNL$\sigma$M
\begin{equation}
    iS_{NLSM} = -\frac{\pi\nu}{8}\Tr\left[D{\left(\hat{\partial}\check{Q}\right)}^2
        + 4 i
        \left(i\hat{\tau}_3 \partial_t \check{Q}  
        + \check{\Delta} \check{Q}\right)
    \right]- i\frac{\nu}{2\lambda}\Tr\check\Delta^\dagger \hat{\gamma}^q \check\Delta.
    \label{eq:nlsm0}
\end{equation}

\subsection{Our system}
We employ a slightly modified NLSM which includes coupling to a thermal bath
\begin{equation}
    iS_{NLSM} = -\frac{\pi\nu}{8}\Tr\left[D{\left(\hat{\partial}\check{Q}\right)}^2
        + 4 i
        \left(i\hat{\tau}_3 \partial_t \check{Q} + i\frac{\gamma}{2}\check{Q}_\text{rel}\check{Q}
        + \check{\Delta} \check{Q}\right)
    \right]- i\frac{\nu}{2\lambda}\Tr\check\Delta^\dagger \hat{\gamma}^q \check\Delta
    \label{eq:nlsm}
\end{equation}
where $D=v_f\tau_\text{imp}^2/2$ is the diffusion constant, $\nu = \nu_\uparrow + \nu_\downarrow$ is the total electronic density of states at the Fermi surface, and $\lambda$ is the strength of the BCS type coupling.
$\Tr$ in the above indicates a trace over all indices: both matrix and spacetime.
The notation $\check{X}$ indicates a matrix in Nambu and Keldysh spaces.
The matrix $\check{Q}$, describing the soft electronic degrees of freedom, is a function of position $\mathbf{r}$ and two time coordinates $t,t'$  and is subject to the non-linear constraint $\check{Q}^2 = \check{1}$.
The photon field $\mathbf{A}$ couples to the model through the covariant derivative
\begin{equation}
    \hat{\partial}\check{X} = \nabla \check{X} + i [\check{\mathbf{A}}, \check{X}]
\end{equation}
where we have absorbed the paramagnetic coupling strength into the definition of the $\mathbf{A}$ field.
All matrices in the model are $4\times 4$ in the product of Keldysh and Nambu spaces.
In what follows we employ the conventions used in Ref.~\onlinecite{kamenev}.
Explicitly
\begin{equation}
\begin{gathered}
    \check{Q}_\text{rel}(\epsilon) =
    \begin{pmatrix}
        1&2F_\text{eq}(\epsilon)\\
        0& -1
    \end{pmatrix}_K\\
    \check{\mathbf{A}} = \sum_\alpha \mathbf{a}_\alpha\hat{\gamma}^\alpha\otimes \hat\tau_3\\
    \check{\Delta} = \sum_\alpha \left(\Delta_\alpha\hat\gamma^\alpha \otimes \hat\tau_+ - \Delta_\alpha^* \hat\gamma^\alpha\otimes \hat\tau_-\right)
\end{gathered}
\end{equation}
where the index $\alpha$ runs over (cl, q) and $\gamma^\text{cl} = \sigma^0$ and $\gamma^\text{q} = \sigma^1$ are matrices in Keldysh space.
We model inelastic relaxation through a linear coupling to a bath $\hat{Q}_\text{rel}$ with temperature $T$ \cite{Tikhonov-2018}.
This is equivalent to the relaxation (1/$\tau$) approximation in the kinetic equation.
In particular $\gamma=\frac{1}{\tau_\text{in}}$ is the inelastic scattering rate.

The saddlepoint equations of Eq.~\eqref{eq:nlsm} for $\Delta^*_q$ and $\check{Q}$ respectively correspond to the BCS gap equation and the Usadel equation\cite{Usadel1970} for the quasiclassical Green's function $\check{Q}$.
In the absence of the cavity photon field this describes the superconducting state of the electronic system without the cavity.
Our strategy will be to obtain the lowest order in $\mathbf{A}$ correction to the action which is linear in $\Delta^*_q$.
This corresponds to the lowest order correction to the gap equation.
In the absence of $\mathbf{A}$ the saddle point of $\check{Q}$ is
\begin{equation}
    \hat{\partial}\left(D \check{Q}\hat{\partial} \check{Q}\right) + i \{i\hat\tau_3\partial_t, \check{Q}\}
    + i\left[i\tau_2\Delta_0 + i\frac{\gamma}{2}\check{Q}_\text{rel}, \check{Q}\right] = 0
\end{equation}
where we have assumed $\Delta_{cl}$ to be homogenous and real.
Assuming a homogeneous, steady state solution $\check{Q}_{sp}(t-t')$ we may Fourier transform to obtain
\begin{equation}
i \epsilon [\hat\tau_3,\check Q(\epsilon)] + i[i\tau_2\Delta_0,\check Q(\epsilon)] + \gamma/2\left[\check{Q}_\text{rel}(\epsilon), \check{Q}(\epsilon)\right] = 0.
\end{equation}
At the saddle point $\check{Q}$ will have the structure
\begin{equation*}
    \check Q =
    \begin{pmatrix}
        \hat{Q}^R& \hat{Q}^R\hat{F} - \hat{F} \hat{Q}^A\\
        0 & \hat{Q}^A
    \end{pmatrix}
\end{equation*}
as governed by fluctuation-dissipation.

\section{Gaussian Fluctuations}
Gaussian fluctuations about the saddle point can be parametrized
\begin{equation}
    \check{Q} = \check{U}\check{V}^{-1} e^{-\check{W}/2} \hat\sigma_3\hat\tau_3 e^{\check{W}/2}\check{V} \check{U}.
    \label{eq:wparam}
\end{equation}
with
\begin{equation}
    \begin{gathered}
    U(\epsilon) =
    \begin{pmatrix}
        1&F_\text{eq}(\epsilon)\\
        0&-1
    \end{pmatrix}_K
    \hat \tau_0\\
    \check{V}(\epsilon) =
    \begin{pmatrix}
        e^{\tau_1 \theta/2}&0\\
        0&e^{\tau_1\theta^*/2}
    \end{pmatrix}_K.
\end{gathered}
\end{equation}
Here, $\theta(\epsilon)$ is a complex angle which is determined by the Usadel equation, and satisfies $\theta(-\epsilon) = -\theta^*(\epsilon)$.
The matrices $U$ and $V$ are a change of basis which allows us to separate the equilibrium and saddle point properties from the fluctuation effects: $U$ describes the fluctuation dissipation relation, while $V$ parametrizes the solution to the retarded Usadel equation.
The matrix $\check{W}$ is then composed of fields multiplying the generators of the algebra which describes rotations on the soft manifold imposed by the nonlinear constraint $\check{Q}^2=1$.
In particular, the matrix $\check{W}$ anticommutes with $\sigma_3\tau_3$ and for $\check{W} = 0$ Eq.~\eqref{eq:wparam} reduces to the saddlepoint solution.
By expanding the exponential in this parametrization we can capture the Gaussian fluctuations along the soft manifold.
$\check{W}$ has 4 independent components that couple to the vector potential
\begin{equation}
    \check{W}(\mathbf r, t, t') =
    i 
    \begin{pmatrix}
        c_R(\mathbf r, t, t')\tau_1& d_{cl}(\mathbf r, t, t') \tau_0\\
        d_q(\mathbf r, t, t') \tau_0& c_A(\mathbf r, t, t')\tau_1
    \end{pmatrix}_K,
\end{equation}
the cooperon ($c_R,c_A$) and diffuson ($d_cl, d_q$) fields.

We now expand Eq.~\eqref{eq:nlsm} to quadratic order in the cooperon and diffuson fields $c$ and $d$.
Doing so we generate three types of terms.
The simplest is the quadratic diffusive mode action
\begin{equation}
  iS_{cd} = \frac{\pi\nu}{4}
  \int\frac{d\epsilon}{2\pi} \int\frac{d\epsilon'}{2\pi}
  \operatorname{tr}
  \left[
    \vec{d}_{\epsilon'\epsilon}
    \hat{\mathcal{D}}^{-1}_{\epsilon\epsilon'}
    \vec{d}_{\epsilon\epsilon'}
    +
    \vec{c}_{\epsilon'\epsilon}
    \hat{\mathcal{C}}^{-1}_{\epsilon\epsilon'}
    \vec{c}_{\epsilon\epsilon'}
    \right]
\end{equation}
where we have defined the vector notation
\begin{equation}
\begin{gathered}
\label{eq:cdvectors}
\vec{d} = (d^{cl}, d^{q})\\
\vec{c} = (c^{R}, c^{A})\\
\hat{D}^{-1}_{\epsilon\epsilon'} = \mathcal{D}^{-1}_{\epsilon'\epsilon}\sigma_{+} + \mathcal{D}^{-1}_{\epsilon\epsilon'}\sigma_{-} \\
\hat{\mathcal{C}}^{-1}_{\epsilon\epsilon'} = \operatorname{diag}\left([\mathcal{C}^{R}_{\epsilon \epsilon'}]^{-1}, [\mathcal{C}^{A}_{\epsilon \epsilon'}]^{-1}\right),
\end{gathered}
\end{equation}
and the diffuson and cooperon propagators
\begin{equation}
\begin{gathered}
\mathcal{D}^{-1}_{\epsilon\epsilon'} = \mathcal{E}^{R}(\epsilon) + \mathcal{E}^{A}(\epsilon')\\
[\mathcal{C}^{R/A}]^{-1}_{\epsilon\epsilon'} = \mathcal{E}^{R/A}(\epsilon) + \mathcal{E}^{R/A}(\epsilon')\\
\mathcal{E}^{R}(\epsilon)  = i\left(\epsilon + i \frac{\gamma}{2}\right)\cosh \theta_{\epsilon} - i \Delta \sinh \theta_{\epsilon}\\
\mathcal{E}^{A}(\epsilon)  = \left(\mathcal{E}^{R}(\epsilon)\right)^{*}.
\end{gathered}
\end{equation}
At linear order we then have a coupling between diffusive modes and the gap
\begin{equation}
  iS_{\Delta-cd} =\pi\nu
\int \frac{d\epsilon}{2\pi} \left[\vec{c}_{\epsilon\epsilon}\cdot \vec{s}^{c}_{\epsilon}  + \vec{d}_{\epsilon\epsilon} \hat{\sigma}_{1}\vec{s}^{d}_{\epsilon}\right]
\end{equation}
where we have taken $\Delta_q$ to be homogeneous and real.
Finally, there is a coupling of the diffusons and cooperons to the photon field
\begin{equation}
\pi\nu  D\int\frac{d\omega}{2\pi} \mathbf{A}^{\alpha}_{-\omega} \cdot \mathbf{A}^{\beta}_{\omega}\int \frac{d\epsilon}{2\pi} \left[\vec{c}_{\epsilon\epsilon}\cdot \vec{r}^{c;\alpha\beta}_{\epsilon}  + \vec{d}_{\epsilon\epsilon} \hat{\sigma}_{1}\vec{r}^{d;\alpha\beta}_{\epsilon}\right]
\end{equation}

The $\vec{r}^{i;\alpha\beta}$ are matrices in the photon Keldysh space and vectors in the sense induced by Eq.~\ref{eq:cdvectors}.  They determined by the structure of the saddlepoint solution and arise from expanding to covariant derivative term in Eq.~\eqref{eq:nlsm} to lowest order in the $W$ matrix fields.

The coupling to the diffusive modes may be removed by making a shift of the fields
\begin{gather}
\vec{c}_{\epsilon\epsilon} \to \vec{c}_{\epsilon\epsilon} - 2 \Delta^{q} \hat{\mathcal{C}}_{\epsilon\epsilon }\vec{s}^{c}_{\epsilon} - 2 D\hat{\mathcal{C}}_{\epsilon\epsilon} \int \frac{d\omega}{2\pi} \mathbf{A}^{\alpha}_{-\omega} \mathbf{A}^{\beta}_{\omega} \vec{r}_{\epsilon}^{c;\alpha\beta }\\
\vec{d}_{\epsilon\epsilon} \to \vec{d}_{\epsilon\epsilon} - 2 \Delta^{q} \hat{\mathcal{D}}_{\epsilon\epsilon}\hat{\sigma_{1}}\vec{s}^{d}_{\epsilon} - 2 D \int \frac{d\omega}{2\pi} \mathbf{A}^{\alpha}_{-\omega} \mathbf{A}^{\beta}_{\omega} \hat{\mathcal{D}}_{\epsilon\epsilon}\hat{\sigma_{1}}\vec{r}_{\epsilon}^{d;\alpha\beta }.
\end{gather}
This shift has three effects.
The first two are to create a nonlinear term in the photon action, which we will ignore as we are not considering non-linear effects, and to create term at second order $\Delta_q$ which we can ignore as $\Delta_q$ will be taken to $0$ at the end.
The important effect is that a coupling between photons and $\Delta_q^*$ is induced
\begin{equation}
iS_{\Delta-A} =
2\pi\nu D \Delta^q \int\frac{d\omega}{2\pi}
    \int \frac{d\mathbf{q}}{(2\pi)^2}\mathbf{A}^{\alpha}_{-\omega}(-\mathbf{q}) \cdot \mathbf{A}^{\beta}_{\omega}(\mathbf{q})
\int \frac{d\epsilon}{2\pi} \left[\vec{s}^{c}_{\epsilon}\hat{\mathcal{C}}_{\epsilon\epsilon}\vec{r}^{c;\alpha\beta}_{\epsilon}
+ \vec{s}^{d}_{\epsilon}\hat{\sigma}_{1}\hat{\mathcal{D}}_{\epsilon\epsilon}\hat{\sigma}_{1}\vec{r}^{d;\alpha\beta}_{\epsilon}\right].
\end{equation}

At this point we may safely integrate out the $d$ modes and henceforth ignore them.\footnote{We are free to ignore the residual coupling to $\Delta$ as the saddlepoint equation guarantees that it vanishes.}

Making the definition
\begin{equation}
-i\Pi^{\alpha\beta} =
2\pi\nu D \Delta^q
\int \frac{d\epsilon}{2\pi} \left[\vec{s}^{c}_{\epsilon}\hat{\mathcal{C}}_{\epsilon\epsilon}\vec{r}^{c;\alpha\beta}_{\epsilon}
+ \vec{s}^{d}_{\epsilon}\hat{\mathcal{D}}_{\epsilon\epsilon}\vec{r}^{d;\alpha\beta}_{\epsilon}\right]
\end{equation}
we can write the photon action as
\begin{equation}
iS_{A} = i\int \frac{d\omega}{2\pi} \int \frac{d\mathbf{q}}{(2\pi)^{2}}\mathbf{A}^{\alpha}_{-\omega, -\mathbf{q}}\left( \check{S}_0^{-1}(\omega, \mathbf{q})
- \check{\Pi}(\omega, \mathbf{q})\right)\mathbf{A}^{\beta}_{\omega, \mathbf{q}}.
\end{equation}
Integrating out $\mathbf{a}$ we obtain
\begin{equation}
iS = -\frac{1}{2}\operatorname{Tr}\ln \left[-i\left(\check{S}^{-1}_{0} - \check{\Pi}\right)\right]
\approx \frac{1}{2}\operatorname{Tr}\left[\check{S}_{0}\check{\Pi}\right]
\end{equation}
where we have expanded to linear order in $\Delta_q$.
Since the momentum $\mathbf{q}$ appears only in $S$ we can immediately integrate over it.
Similarly we can trace over the in plane components of A.
We thus define
\begin{equation}
\hat{D}(\omega) = \sum_{i \in \{x, y\}} \int \frac{d\mathbf{q}}{(2\pi)^2} \hat{S}^{ii}(\omega, \mathbf{q})
\end{equation}
We assume the photon modes to be governed by a density matrix which is diagonal in energy.
$D$ can then be written in the usual form
\begin{equation}
\hat{D}(\omega) =
\begin{pmatrix}
N(\omega)(D^R(\omega) - D^A(\omega))& D^R(\omega)\\
D^A(\omega)&0
\end{pmatrix}
\end{equation}
Defining $-2\pi i J(\omega) = D^R(\omega) - D^A(\omega)$ and using the analytic properties of $D$ this can be written
\begin{equation}
iS = \frac{-i}{2}\int d\omega J(\omega)\left[N(\omega)\Pi_{0, 0}(\omega)
- (\Pi^R(\omega) - \Pi^A(\omega))\right]
\end{equation}
where we have defined $\Pi^{R/A}$ as the retarded/analytic part of $\Pi^{01/10}$.
Defining 
\begin{gather}
\nu \Delta^{q} \left(P_{\alpha\beta}^c(\omega)
+ P_{\alpha\beta}^d(\omega) \right)
= -i\Pi^{\alpha\beta}\\
\mathcal{B}(\omega) = \frac{P^d_R(\omega) - P^d_A(\omega)}{P^d_0(\omega)}
\end{gather}
with $P^0 = P_{00}$  and $P^{R/A}$ defined analogously to $\Pi^{R/A}$ the correction can be broken into two terms.
The first is the equilibrium self-energy correction to to the cavity photons
\begin{equation}
    iS_{c}^{eq} = \frac{\nu\Delta^q}{2}\int d\omega J(\omega)\left[\mathcal{B}(\omega)P^{c}_{0}(\omega)
- (P^{c}_{R}(\omega) - P^{c}_A(\omega))\right].
\end{equation}
This term should be included in the bare equilibrium result as it is a property of the equilbrium cavity-superconductor system and we therefore subtract it off henceforth.
The other term 
\begin{equation}
iS_{fluc} = \frac{\nu\Delta^q}{2}\int d\omega J(\omega)(N(\omega) - \mathcal{B}(\omega))(P^{c}_{0}(\omega) + P^{d}_{0}(\omega))
\end{equation}
is the fluctuation induced enhancement to superconductivity.
This is to be compared with the correction term due to a classical monochromatic field (i.e. the original Eliashberg effect)
\begin{equation}
iS = (-i\Pi_{0,0}(\omega) -i\Pi_{0,0}(-\omega))|\mathbf{A}_{\omega}|^2
= \nu\Delta^{q}(P_{0}(\omega) + P_{0}(-\omega))|\mathbf{A}_{\omega}|^2
\equiv \nu \Delta^q Y(\omega)|\mathbf{A}_{\omega}|^{2}.
\end{equation}
Using the functional dependence of the classical Eliashberg effect on frequency $Y(\omega)$ the quantum Eliashberg effect can be written in a Fluctuation-Dissipation like form
\begin{equation}
iS^{\mathrm{fluc}} = \frac{\nu\Delta^q}{2}\int_0^{\infty} d\omega J(\omega)(N(\omega) - \mathcal{B}(\omega))Y(\omega)
\label{eq:linear_correction}.
\end{equation}
It should be noted that in the linearized regime $P_0^d$ goes as $\gamma^{-1}$ while $P_0^c$ goes as $\gamma^0$.
Thus, in the limit of $\gamma\to0$ we expect the diffuson contribution to be dominant.

\section{Gap Equation}
As mentioned previously, the BCS gap equation is the saddlepoint equation of our action with respect to the source field $\Delta_q$.
Including the correction term Eq.~\eqref{eq:linear_correction} the gap equation then becomes
\begin{equation}
    0 = \left.\frac{\delta i S}{\delta \Delta_q}\right|_{\Delta_q = 0} =
    - 4 i \frac{\nu}{\lambda} \Delta + \frac{\pi\nu}{2} \Tr\hat{Q}^K\hat\tau_2
+  \frac{\nu}{2}\int_0^{\infty} d\omega J(\omega)(N(\omega) - \mathcal{B}(\omega))Y(\omega)
\end{equation}
We therefore define
\begin{equation}
    \begin{gathered}
    F_\text{BCS} = \frac{1}{\lambda} + \frac{i\pi}{8\Delta} \Tr\hat{Q}^K\hat\tau_2\\
    F_\text{phot} = 
\frac{i\nu}{8\Delta}\int_0^{\infty} d\omega J(\omega)(N(\omega) - \mathcal{B}(\omega))Y(\omega)
    \end{gathered}
\end{equation}
Which allows us to write the gap equation as $F_\text{BCS} = - F_\text{phot}$.
Furthermore, $F_\text{phot}$ can be broken up into a kinetic contribution $F^\text{kin}$ arising from modification of the quasiparticle occupation function and a spectral contribution $F^\text{spec}$ due to modification of the density of states from self energy effects, as discussed above.
Most notably, because the gap equation is linearly related to the action, the corrections to the gap equation are related to the conventional via the same fluctuation-dissipation-like relation.

\subsection{Effective photonic spectral function}

The function $J(\omega)$ can be can be calculated by relating the field $\mathbf{A}$ to the cavity mode operators $a,\bar{a}$.

\subsubsection{Multimode Cavity}
As an example of a multimode cavity we take the cavity mode Keldysh action to be given by
\begin{equation}
iS = i\int \frac{d\omega}{2\pi}\int \frac{d\mathbf{q}}{(2\pi)^2}
a^{\dagger}_{q;\alpha}
\underbrace{
\begin{pmatrix}
0 & \omega - i\kappa - \omega_q\\
\omega + i\kappa - \omega_q&2i\kappa N(\omega)
\end{pmatrix}
}_{\hat{G}^{-1}(\omega, \mathbf{q})}
a_{q;\alpha}.
\end{equation}
to describe a cavity coupled to the environment.\cite{Diehl-2016}
Using the fact that we can expression $\mathbf{A}$ in terms of $a$ and $\bar a$ (in Gaussian units) as
\begin{gather}
\mathbf{A}_{q}(z) = \sqrt{\frac{2\pi c^{2}}{\omega_{q}}}\left(a_{q;\alpha} \epsilon_{\mathbf{q};\alpha}(z) + a^{\dagger}_{-q;\alpha} \epsilon^{*}_{-\mathbf{q};\alpha}(z) \right)
\end{gather}
we can relate the Keldysh component of $S$ and $G$
\begin{equation}
2S^{K}_{\omega, \mathbf{q}; ii}(L/2, L/2) 
= \frac{2\pi c^{2}}{\omega_{q}}
\sum_{\alpha}
|\epsilon^{i}_{\mathbf{q};\alpha}(L/2)|^{2}
\left(
G^{K}_{-q} + G^{K}_{q}
\right)
\end{equation}
After some calculation we therefore find
\begin{equation}
J_\text{MM}(\omega) =
\int \frac{d\mathbf{q}}{(2\pi)^2}
\frac{\kappa c^{2}}{\omega_{q}}
\sum_{\alpha}\left|\epsilon_{\mathbf{q};\alpha}\left(\frac{L}{2}\right)\right|^{2}
\left(
\frac{1}{(\omega - \omega_{q})^{2} + \kappa^{2}}
 - \frac{1}{(\omega + \omega_{q})^{2} + \kappa^{2}}
\right)
\end{equation}
where we have used the fact that $\epsilon(L/2)$ is in plane.
Now with the explicit forms of $\epsilon_i$ from the main text
\begin{equation}
    \begin{gathered}
  \hat{\epsilon}_{1,\mathbf{q}}(L/2) = -i\sqrt{\frac{2}{L}} \frac{\omega_0}{\omega_\mathbf{q}} \frac{\mathbf{q}}{|\mathbf{q}|}\\
  \hat{\epsilon}_{2,\mathbf{q}}(L/2) = \sqrt{\frac{2}{L}} \mathbf{e}_3 \times \frac{\mathbf{q}}{|\mathbf{q}|}
  \end{gathered}
\end{equation}
we can immediately evaluate the angular integral
\begin{equation}
    \int \frac{d\theta}{2\pi} \sum_{i\in{x, y}, \alpha}|\epsilon^i_{\theta, \alpha}(L/2)|^2 = \frac{2}{L}\left(1 + \frac{\omega_0^2}{\omega_{\mathbf q}^2}\right).
\end{equation}
We now make a change of variables from $|\mathbf{q}|\to\omega' = \omega_{\mathbf q}$.
The dispersion relation $\omega_q^2 = \omega_0^2 + c^2 q^2$ implies
\begin{equation}
    \frac{q dq}{2\pi \omega'} = \frac{d\omega'}{2\pi c^2}.
\end{equation}
This allows us to write $J$ as
\begin{equation}
    J_\text{MM}(\omega) = \frac{2\kappa}{L} \int_{\omega_0}^\infty d\omega' 
\left(\frac{1}{(\omega - \omega')^2 + \kappa^2} - \frac{1}{(\omega + \omega')^2 + \kappa^2}\right)
\left(1 + \frac{\omega_0^2}{\omega'^2}\right).
\end{equation}
This integral may be performed exactly to find
\begin{multline}
    J_\text{MM}(\omega) = 
    \frac{2}{L}
    \left[
    \left(
    1 + \omega_0^2\frac{\omega^2 - \kappa^2}{(\omega^2 + \kappa^2)^  2}\right)
    \left(
    \tan^{-1}\frac{\omega - \omega_0}{\kappa}
    + \tan^{-1}\frac{\omega + \omega_0}{\kappa}
    \right)\right.\\
    \left.
    + \frac{\kappa\omega\omega_0^2}{\left(\omega^2 + \kappa^2\right)^2}
    \log\left(
    \frac{\left((\omega -\omega_0)^2 + \kappa^2\right)\left((\omega +\omega_0)^2 + \kappa^2\right)}{\omega_0^4}
    \right)
    \right].
\end{multline}
We will, however, introduce a factor $X$ into $J$ which describes enhancement of the electron-photon coupling due to e.g. squeezing of mode volume, one factor of $\sqrt{X}$ coming from the enhancement of each vertex.
In principle this enhancement should come from a detailed study of the structure of the photon modes.
However, this physics is not captured within our simple parallel plate model and so we include the coupling enhancement phenomenonlogically via the factor $X$
\begin{equation}
    J_\text{eff}(\omega) = X J(\omega).
\end{equation}

\subsubsection{Single mode cavity}
We can also consider the effective photonic spectral function for a single mode cavity
\begin{equation}
iS = i\int \frac{d\omega}{2\pi}
a^{\dagger}_{\alpha}(\omega)
\underbrace{
\begin{pmatrix}
0 & \omega - i\kappa - \omega_0\\
\omega + i\kappa - \omega_0&2i\kappa N(\omega)
\end{pmatrix}
}_{\hat{G}^{-1}(\omega)}
a_{\alpha}(\omega).
\end{equation}
Following the steps outlined above we find that
\begin{equation}
    J_\text{eff;SM}(\omega) = 
\frac{\kappa c^{2}X}{ \omega_{0}}
\sum_{\alpha}\left|\epsilon_{\alpha}\left(\frac{L}{2}\right)\right|^{2}
\left(
\frac{1}{(\omega - \omega_{0})^{2} + \kappa^{2}}
 - \frac{1}{(\omega + \omega_{0})^{2} + \kappa^{2}}
\right).
\end{equation}

\subsection{Photonic corrections to the distribution function}

To lowest order in $\tau_\text{in}=1/\gamma$, which corresponds to taking a linearized expansion of the collision integral in the deviation of the occupation function from Fermi-Dirac,
and using the fact that $J(\omega)$ is an odd function of $\omega$ we can write $F^\text{kin}_\text{phot} = F_\text{pair} + F_\text{scat}$ with the recombination contribution
\begin{equation}
    F_\text{pair}=\frac{\alpha D}{\gamma c} \int_{2 \Delta}^{\infty}d\omega J\left(\omega\right)\left(N{\left (\omega \right )} - \mathcal{B}{\left (\omega \right )}\right)
 \int_{\Delta}^{\omega - \Delta} \frac{d\epsilon}{\epsilon} \left(F{\left (\epsilon \right )} + F{\left (\omega - \epsilon \right )}\right)
 P{\left (\epsilon,\omega - \epsilon \right )}  \rho_{qp}{\left (\epsilon \right )} \rho_{qp}{\left (\omega - \epsilon \right )}
\end{equation}
and scattering contribution
\begin{equation}
    F_\text{scatter}=  \frac{\alpha D}{\gamma c} \int_{0}^{\infty}d\omega\,\omega J\left(\omega\right)\left(N{\left (\omega \right )} - \mathcal{B}{\left (\omega \right )}\right)
    \int_{\Delta}^{\infty} \frac{d\epsilon}{\epsilon(\epsilon+\omega)} \left(F{\left (\epsilon \right )} - F{\left (\omega + \epsilon \right )}\right)
 L{\left (\epsilon,\omega + \epsilon \right )}  \rho_{qp}{\left (\epsilon \right )} \rho_{qp}{\left (\omega + \epsilon \right )},
\end{equation}
where the fine-structure constant $\alpha$ appears due to reinstating the electron charge in the paramagnetic coupling which we had previously absorbed into the $\mathbf{A}$ field.

With our particular form of $J(\omega)$ ($G(w, k)$) the correction to the gap equation become
\begin{equation}
    F_\text{pair}=\frac{\alpha DX}{c\gamma } \int_{2 \Delta}^{\infty}d\omega J\left(\omega\right)\left(N{\left (\omega \right )} - \mathcal{B}{\left (\omega \right )}\right)
 \int_{\Delta}^{\omega - \Delta} \frac{d\epsilon}{\epsilon} \left(F{\left (\epsilon \right )} + F{\left (\omega - \epsilon \right )}\right)
 P{\left (\epsilon,\omega - \epsilon \right )}  \rho_{qp}{\left (\epsilon \right )} \rho_{qp}{\left (\omega - \epsilon \right )}
\end{equation}
and
\begin{equation}
    F_\text{scatter}=\frac{\alpha DX}{c\gamma }  \int_{0}^{\infty}d\omega\,\omega J\left(\omega\right)\left(N{\left (\omega \right )} - \mathcal{B}{\left (\omega \right )}\right)
    \int_{\Delta}^{\infty} \frac{d\epsilon}{\epsilon(\epsilon+\omega)} \left(F{\left (\epsilon \right )} - F{\left (\omega + \epsilon \right )}\right)
 L{\left (\epsilon,\omega + \epsilon \right )}  \rho_{qp}{\left (\epsilon \right )} \rho_{qp}{\left (\omega + \epsilon \right )}.
\end{equation}
In the above we have used the definitions
\begin{equation}
    \begin{gathered}
        P(\epsilon, \epsilon') = 1 - \frac{\Delta^2}{\epsilon\epsilon'},\qquad
        L(\epsilon, \epsilon') = 1 + \frac{\Delta^2}{\epsilon\epsilon'}\\
        F(\epsilon) = \tanh \frac{\epsilon}{2T},
        \qquad
        N(\omega) = \coth \frac{\omega}{2T_p},
        \qquad
        \mathcal{B}(\omega) = \coth \frac{\omega}{2T}
    \end{gathered}
\end{equation}
We have assumed the photons to be at temperature $T_p$ while the Fermions are coupled to a bath of temperature $T$.

The correction terms can be rewritten as
\begin{multline}
F_\text{pair} + F_\text{scat} =\frac{\alpha DX}{\gamma c } \int_\Delta^\infty d\epsilon \frac{\rho_{qp}(\epsilon)}{\epsilon} \int_0^{\infty}d\omega J(\omega) \left(N{\left (\omega \right )} - \mathcal{B}{\left (\omega \right)} \right)\\
 \times  \left[\left(F{\left (\epsilon \right )} + F{\left (\omega - \epsilon \right )}\right)
 P{\left (\epsilon,\omega - \epsilon \right )} \rho_{qp}{\left (\omega - \epsilon \right )}\Theta(\epsilon - \Delta)\Theta(\omega - \Delta - \epsilon)\right.\\
 + \left(F(\epsilon) - F(\epsilon + \omega)\right)L(\epsilon, \epsilon + \omega)\rho_\text{qp}(\epsilon + \omega)\Theta(\epsilon - \Delta)\\
 \left.
 + \left(F(\epsilon - \omega ) - F(\epsilon)\right)L(\epsilon - \omega, \epsilon)\rho_\text{qp}(\epsilon - \omega)\Theta(\epsilon - \omega - \Delta)
 \right]
 = 2\int_\Delta^\infty d\epsilon \frac{\rho_{qp}(\epsilon)}{\epsilon} n_1(\epsilon)
\end{multline}
which allows us to move this term to the left hand side to obtain
\begin{equation}
    \frac{1}{\lambda} - \int_{\Delta}^{\infty} d\epsilon \frac{1 - 2n_f(\epsilon) - 2n_1(\epsilon)}{\sqrt{\epsilon^2- \Delta^2 }}=0
\end{equation}
and therefore identify the correction to the occupation function
\begin{multline}
    n_1
=\frac{\alpha DX}{2\gamma c }  \int_0^{\infty}d\omega J(\omega) \left(N{\left (\omega \right )} - \mathcal{B}{\left (\omega \right)} \right)\\
 \times  \left[\left(F{\left (\epsilon \right )} + F{\left (\omega - \epsilon \right )}\right)
 P{\left (\epsilon,\omega - \epsilon \right )} \rho_{qp}{\left (\omega - \epsilon \right )}\Theta(\epsilon - \Delta)\Theta(\omega - \Delta - \epsilon)\right.\\
 + \left(F(\epsilon) - F(\epsilon + \omega)\right)L(\epsilon, \epsilon + \omega)\rho_\text{qp}(\epsilon + \omega)\Theta(\epsilon - \Delta)\\
 \left.
 + \left(F(\epsilon - \omega ) - F(\epsilon)\right)L(\epsilon - \omega, \epsilon)\rho_\text{qp}(\epsilon - \omega)\Theta(\epsilon - \omega - \Delta)
 \right].
\end{multline}
Defining the power spectral density of absorption
$(\alpha D/c) J(\omega),$
our result can be written
\begin{equation}
    n_1(\epsilon) = \gamma^{-1}\int_0^\infty d\omega\ S(\omega)\frac{N(\omega) - \mathcal{B}(\omega)}{2} I_\epsilon^\text{el}(\omega)
\end{equation}
where $I^\text{el}_\text{eps}(\omega)$ is the related to the conventional Eliashberg expression~\cite{Eliashberg-1973} for a classical microwave field $\mathbf{A}_\omega$ by
\begin{equation}
    n^\text{conv.}_1(\epsilon, \omega) =
    \frac{\alpha D |\mathbf{A}_\omega|^2}{\gamma c}I^\text{el}_\epsilon(\omega).
\end{equation}




\end{document}